# Visualizing the invisible – the relentless pursuit of MTech Imaging

Teaching Case

**Jenson Chong-Leng Goh**
School of Business
SIM University
Singapore
Email: jensongohcl@unisim.edu.sg

**Jeffery Beng-Huat Tan**
SIM Professional Development
Singapore Institute of Management
Singapore
Email: jtan@sim.edu.sg

**Janice Sio-Nee Tan**
SIM Professional Development
Singapore Institute of Management
Singapore
Email: janicetansn@sim.edu.sg

## Abstract

This teaching case describes the challenges faced by MTech Imaging, a Singapore small and medium enterprise (SME) that specializes in providing thermal imaging solutions. In recent years, the company has relentlessly strived to become a digital innovative solution provider. This push has led to the development of a disruptive digital innovation called the AXION platform. Students are provided with vivid accounts of the journey undertaken by MTech to develop the AXION platform, the industry it competes in, and the challenges it faces in attempting to disrupt its industry through the introduction of the AXION platform. The case seeks to achieve three learning objectives: (1) allow students to learn from MTech's experiences in developing disruptive digital innovation; (2) immerse students as senior management of MTech to substantiate the best digital innovation strategy to adopt in order to disrupt its industry; and (3) expose students to the challenges of driving the adoption of such digital innovation in the market. It is hoped that the case can inspire students to become effective digital innovation entrepreneurs.

# 1   Introduction

It was a warm morning in Singapore.

But it was not as 'warm' as the intense discussion in the MTech Imaging Pte Ltd (MTech) office located at the Changi Business Park. Three senior executives responsible for the future of the company were debating and making bold plans to make history in the thermal imaging industry. The room was filled with excitement and anxieties as the discussion heated up.

"*I think we are creeping into a blue ocean market when we make this technology available to our customers who often say that if a company is to come up with such technology, they will definitely buy it.*" – Mr Richard Yong, Chief Operating Officer, Miltrade Holdings Pte Ltd.

Richard is the Chief Operating Officer of Miltrade Holdings Pte Ltd, the parent company of MTech. The technology that Richard referred to is MTech's ground-breaking open camera platform, AXION (see Exhibit 1). Conceived and developed entirely by MTech's in-house research and development team, this platform is a fully programmable and customizable infrared processing platform that incorporates a custom optical design, multi-brand image fusion, non-optical sensors, and advanced computational imaging techniques and applications to create extremely high quality infrared images that were previously unseen in the thermal imaging industry (see Exhibit 1 for details).

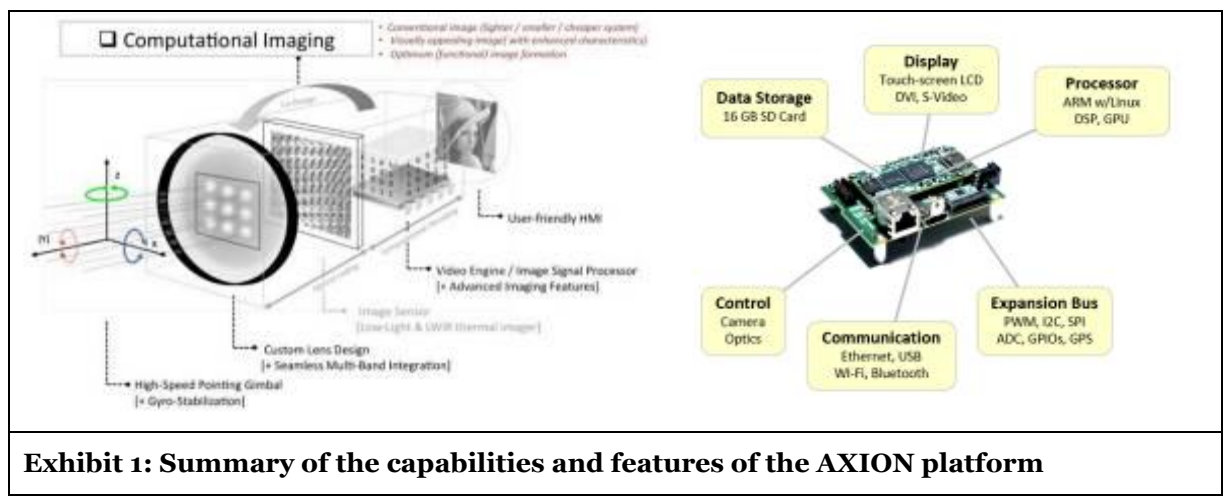

**Exhibit 1: Summary of the capabilities and features of the AXION platform**

"*I believe law enforcement is going to be our big customers because AXION is going to be light and cheap enough to be installed onto a handheld, which allows law enforcers to take it out into the field.*" – Mr Bob Nishi, General Manager, MTech Imaging Pte Ltd.

"*Bob, we need to think about what we can offer to our customers first. I think we need a platform where we allow researchers and programmers to sell their applications and algorithms [in thermal imaging processing] to us and then we integrate them into our AXION technology. With sufficient contributions from these communities, we will be able to create a repertoire of software libraries that will make AXION extremely attractive to any customers. Then, we can determine the customer segments to target based on these contributions*" – Mr Baljit Singh, Executive Director, MTech Imaging Pte Ltd.

Since the mid of 2013, the meetings among the three senior executives namely, Richard, Baljit and Bob, had been dominated by this relentless push and heated debate to fulfil the full potentials of its AXION platform. As a SME[1], the company has limited financial and human resources. The three senior executives had been debating passionately about the various strategic moves that can provide the maximum benefits to MTech using the least resources, but they were unable to settle on the solution.

---

[1] According to SPRING Singapore, a SME in Singapore is defined as any company that has either an annual sales turnover of not more than S$100 million or it's employment size does not exceed 200 workers. Under this definition, MTech Imaging is classified as a SME. See details on
http://www.spring.gov.sg/NewsEvents/PR/Documents/Fact_Sheet_on_New_SME_Definition.pdf
and https://www.linkedin.com/company/mtech-imaging-pte-ltd

Suddenly, there was an awkward silence in the warm meeting room.

Richard went into a deep thought mode as he considered carefully all the alternatives proposed during the heated debates. On one hand, he wanted the AXION platform to bring his company to the next level of performance and transform the entire thermal imaging industry. To do that, he knew that time is needed to build the right capabilities. On the other hand, he knew that time was his greatest enemy. The longer he waited, the higher the chance that his competitors (especially the market incumbent) may get wind of his company's invention and possibly catch up.

*"How can we bring our technology [AXION] to our customers in the fastest and most cost-effective way? … Team, I think the solution may be in …"* – Mr Richard Yong, Chief Operating Officer, Miltrade Holdings Pte Ltd.

## 2　Organization Background

MTech is a Singapore-incorporated thermal imaging company that specializes in the design, manufacturing, and sales of thermal imaging products, solutions, and services. The company is a subsidiary of Miltrade Group.

Miltrade Group (Miltrade) was incorporated in the late 1980s and grown into a group of companies that provide innovative technologies and engineering solutions to the Safety and Security industries in Asia. It operates through its subsidiaries, affiliates, and representative offices located in Singapore, Vietnam and India. It engages in technology/product innovation, provisioning of integrated solutions and maintenance/repair and overhaul (MRO) business activities to service the Aerospace, Defence, and Imaging business sectors in Asia. The company invested heavily in its infrastructure in Singapore and Vietnam to support these activities. Given its dependency on technological innovations to drive growth, the Group has an unwavering belief on the recruitment, retention, and management of talents and the development and management of strategic partnerships. Two fundamental principles that the Group adheres to are: (1) the investment in the recruitment and development of talents that are highly competent in driving its business strategies; and (2) the forging, strengthening and expanding of alliances with established strategic partners in different domains.

**Miltrade Group**

| Miltrade Technologies Pte Ltd (Miltrade Technologies) | MTech Imaging Pte Ltd (MTech) | MTech Imaging USA LLC (MTech USA) | Vietstar Aviation Services Joint Stock Company (VAS) | HML Pte Ltd |
|---|---|---|---|---|
| Incorporated in 1993. Provides integrated solutions and maintenance, repair & overhaul (MRO) services to naval & air force customers both locally in Singapore and regionally. Also provides similar services/products to commercial aircraft operators and MRO companies. | Incorporated in 2004. Imaging product system integrator and technology and innovative solutions provider in thermal imaging for the Asian emerging market. In-house R&D team with proven track records to develop innovative imaging products. | Incorporated in 2009 via a joint venture with AugenTek LCC of Dallas, Texas. Provides thermal imaging and night vision solutions to defence customers in both US and overseas. Collaborates closely with MTech Imaging Pte Ltd in Singapore. | Incorporated in 2010 via a joint venture with local partners in Ho Chi Minh, Vietnam. Provides integrated airframe and component MRO services for aircrafts. Offers material sales and management and line maintenance services for commercial aviation operations in Tan Son Nhat Airport and other stations in Vietnam. | Incorporated in 2010 via a joint venture between Miltrade Technologies and an established OEM battery partner. Specializes in the design and assembly of high performance and high reliability Nickel Cadmium Sintered Plate (NSCP) batteries for the commercial aviation market. |

**Exhibit 2: Overview of Companies under Miltrade Group**

The Group is also focused on sensing of trends in its industry and the needs of its customers. Having highly experienced senior management with a combined total of more than five decades, significant amount of their time involved scanning the competitive environment to keep pace with the industry and needs of its customers.

The idea to start up MTech was mooted during Singapore's SARS crisis in 2003 when the Group was called upon by the Singapore Government to quickly assemble thermal imaging scanners that can be deployed at airports and custom points to detect people with fever. This was a vote of confidence and trust from the Singapore Government on the Group's innovative and development capabilities despite the extremely daunting task for a SME given the tight deadline to deliver the scanners in large quantities.

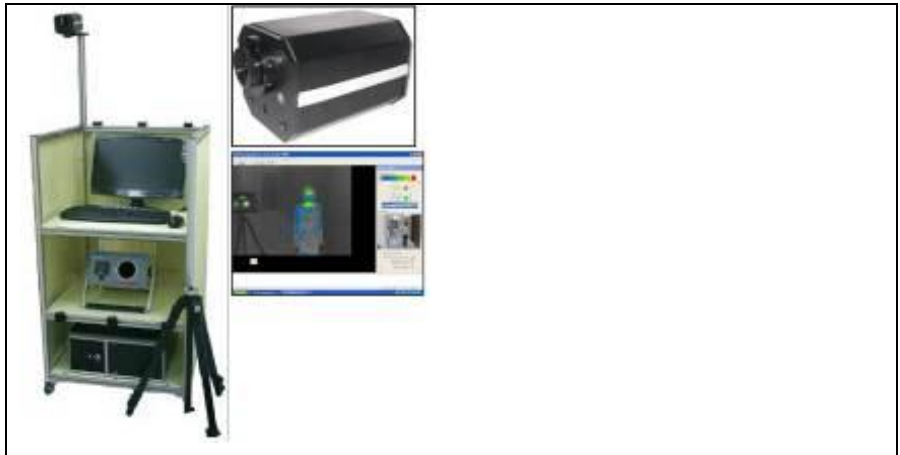

**Exhibit 3: Latest FeverScan S3000 model by MTech Imaging**

Baljit recalled how the company went about meeting the Government's requirements.

*"I had a chance to work with academics … the algorithms were written by an academic but the domain subject expertise came from doctors … who advised on how to correlate body temperature with fever and what you would accept as reliable and unreliable readings."* – Mr Baljit Singh, Executive Director, MTech Imaging Pte Ltd

The Group's research and development team sprung into action immediately. Leveraging its networks and partners, the Group was able to deliver and supply a large number of high quality fever scanners at the height of the SARS crisis. It was an impressive feat in the Group's history which had CNN reported the scanner as one of Singapore's five tech inventions that rocked our world[2]. The episode also made the Group realized there was a large market in the thermal imaging industry for system integration. Having proven itself during the SAR crisis, MTech Imaging Pte Ltd (MTech) was incorporated in 2004 to capture this market opportunity.

MTech started out as an imaging product system integrator focusing on the assembly of existing thermal imaging technologies into customized products. Given the close connection with its parent company, MTech was granted direct access to the defence and aerospace industries where its competencies in thermal imaging were highly applicable and relevant. This allowed the company to penetrate these traditionally high entry barrier sectors, and grew its businesses rapidly. Over time, the company established an extensive sales, marketing, and distribution network in South East Asia, South Asia, and North Asia. Operations was also established in South Africa with plans for markets in the Middle East and Oceania. This extensive network is one of MTech's key strengths which enabled them to bring existing and new products to the market quickly. As of 2014, the MTech's is organized as two business divisions namely, radiometric and non-radiometric, with over five lines of products that cater to three major market segments namely, industrial, safety and security, and military. In the radiometric division, the products are mainly used in oil and gas, power, energy, marine, semiconductor, manufacturing industries, and by commercial institutions in the research and medical fields. The products include preventative maintenance cameras and fever scanners. In the non-radiometric division, the two dominant markets are the military and law enforcement, and safety and security. The products include the Hotspot lightweight thermal weapon sight, Night Scan night vision security camera, and driving aid called DriverViewer (see Exhibit 4 for details of MTech's capabilities & product offerings).

---

[2] Larry, L. (2010). "5 Singapore tech inventions that rocked our world." Retrieved 7th Oct, 2014, from http://travel.cnn.com/singapore/shop/5-best-tech-inventions-singapore-rocked-our-world-423291.

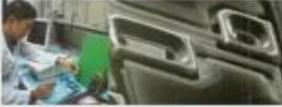
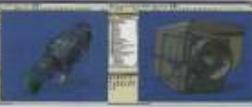

**Exhibit 4: Summary of the current capabilities & product offerings of MTech**

## 3   The Thermal Imaging Industry's Value Chain

The value chain in the thermal imaging industry consists of three key segments, namely, system integrators, finished product distributors, and detector manufacturers (see Exhibit 5 for details of the Thermal Imaging Industry's value chain).

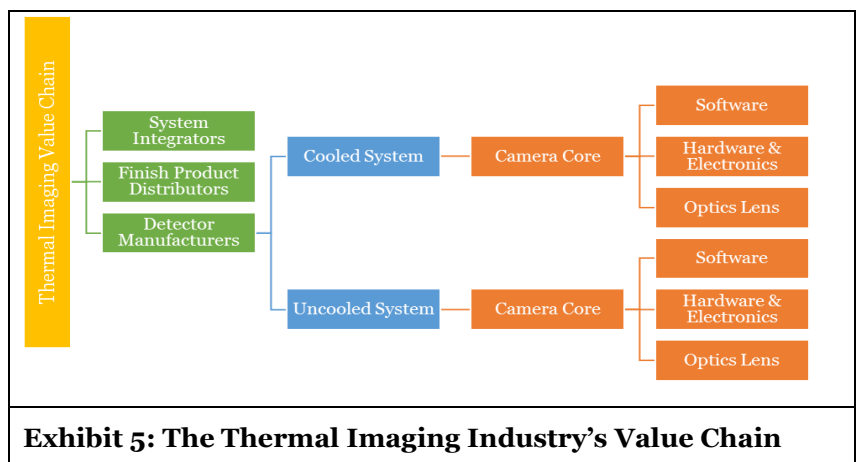

**Exhibit 5: The Thermal Imaging Industry's Value Chain**

System integrators segment are very much like contractors in the building and construction industry. They earn a premium from their customers by sourcing parts and technologies to custom build thermal imaging solution to meet specific requirements. Some examples include specialized thermal imaging devices used in automobiles (e.g. BMW), aircrafts, and medical equipment. Finished product distributors are responsible for sales and distribution of ready-made thermal imaging products for the mass market. These companies typically do not engage in any research and development activities. They are oriented towards providing standardized solutions to the mass market. Detector manufacturers are responsible for manufacturing thermal imaging products or its parts which are often used by system integrators and/or distributed by finished product distributors. They are the industry product innovators that cater to the needs of both the system integrators and finished product distributors.

Detector manufacturers can be divided into two group of companies namely, cooled systems and uncooled systems, defined by the type of thermal imaging detectors that are being manufactured. Detector manufacturers are further supported by a diverse group of the companies that design, develop, and manufacture cameras for use in specific thermal imaging applications. Three key types of support companies include:

1. Software companies that write software to enable or maximize the thermal imaging capabilities of a camera core;
2. Electronics companies that develop the electronics and hardware within the camera core; and
3. Optics lens companies that design and build the specialized lenses used in the camera.

The major market players in the industry are FLIR (www.flir.com – around 40%), DRS Technologies (www.drsinfrared.com – around 20%), and ULIS (www.ulis-ir.com) who are mainly detector manufacturers and finished product distributors. Together, these three companies command more than 60% market share of the thermal imaging industry. Some companies compete exclusively within a specific segment in the value chain. For instance, MTech competes initially only in the system integrators segment while FLIR competes across multiple segments of finished product distributors and detector manufacturers[3]. The competition in the system integrators market has intensified with the commoditization of advanced thermal imaging technologies, which dilutes the value proposition to customers.

*"There are iPhone accessories that have a thermal imaging capability built into it. Not as high resolution as the one we have [in the AXION platform], but it is going to be available for USD350 or so and people is going to be able to snap it on to their iPhone 5 like a case and view thermal imagery [on it] … probably there will be Apps on the AppStore developed specifically for thermal imaging"* – Mr Bob Nishi, General Manager, MTech Imaging Pte Ltd

## 4 A Competitive Move

Management of MTech felt the pains and frustrations in meeting customized needs of their customers due to limitations of existing technologies. Traditionally, it had to work very closely with finished product distributors and detectors manufacturers to design, assemble, and develop a comprehensive solution needed by customers. However, such endeavour is finance, resources and time intensive. Exhibit 6 illustrates the process of developing customized solutions from the initial camera core design to the finished product.

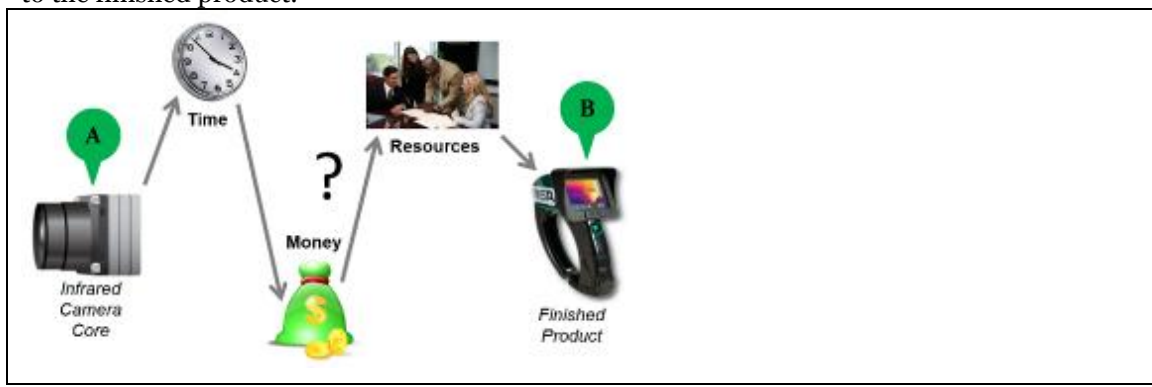

---

[3] A detailed analysis of the thermal infrared market is not within the scope of this document. However, you should be able to find relevant documents on the web (e.g. http://www.prnewswire.com/news-releases/infrared-imaging-market-thermal-imaging-systems-industry-analysis-and-forecasts-498401411.html).

**Exhibit 6: Challenges faced by System Integrators in Developing Finished Product**

Many system integrators are also facing similar challenges. This convinced MTech's management on the potential gap in the industry that can be fulfilled through the development of an integrated turnkey solution like the AXION platform. This is best illustrated in Exhibit 7.

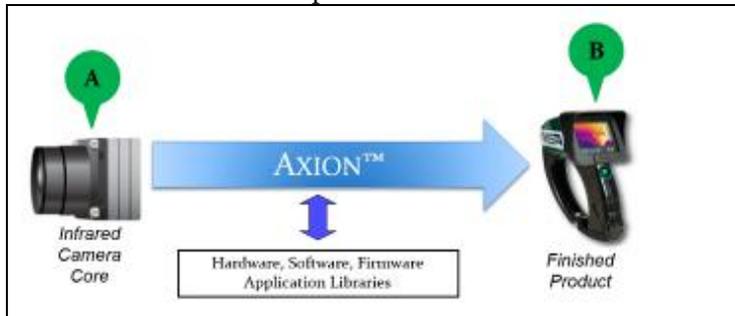

**Exhibit 7: How AXION fills this Gap in the Industry**

The AXION platform reduces the need to spend significant amount of efforts, time and resources to develop, test, and produce the hardware to make the thermal sensor smart and useable for respective target market segments. The AXION platform offers versatility in meeting diverse customer's requirements rapidly.

*"How to get from point A to point B? The AXION platform provides the most cost-effective path to bridge gap between the infrared camera core and the finished product. Within the AXION platform is the hardware, software and application libraries that host the algorithms to do the analysis for [infrared] image enhancements so that the [infrared] image you see in the finished product would be of the best quality"* – Mr Bob Nishi, General Manager, MTech Imaging Pte Ltd

The development of the AXION platform allowed MTech to successfully transform from just being a system integrator to a product innovator in the industry. For instance, the detector manufacturers can make use of the 'plug-and-play' flexibility in the AXION platform to shorten the development time needed to develop customized finished products for its customers. This could potentially render the entire camera core companies irrelevant and weaken the value proposition of companies providing customized thermal imaging solutions. Similarly, finished product distributors could also be significantly impacted as cost of developing customized thermal imaging products is one of the reasons mass market customers are inclined to purchase finished products. With this cost significantly reduced, customers who initially chose finished products could now consider the system integration option.

Given that system integrators do not typically invest in the research and development of thermal imaging technologies, there is a potentially large demand looking for an efficient and cost-effective solution provided by the AXION platform. By using the AXION platform, customized products can be developed quicker and at a faction of cost that was incurred in the past[4].

## 5 Making the Move

### 5.1 Assembling the AXION Team

Convinced this was the way to create a competitive advantage, MTech's management started a search in 2009 for relevant talents to help them fulfil their dreams. This search was extremely challenging because there wasn't a critical pool of thermal imaging talents in Singapore. Given this domestic constraint, talent sourcing shifted overseas where there was a thriving community of experts especially USA.

*"When we first assembled this team, Bob would tell you that we had 10 super duper engineers and scientists [from overseas] and none of them has thermal imaging experiences. It took us some time to*

---

[4] Example of the products that can be customized through the use of the AXION platform can be found in this URL: http://miltrade.com.sg/index.php/imaging-business/imaging-applications/imaging-axion

*develop this AXION. We simply could never find such expertise here [in Singapore]"* – Mr Baljit Singh, Executive Director, MTech Imaging Pte Ltd

*"It was not easy to assemble the engineering team for AXION. We had countless CVs to screen and had conducted at least two rounds of interviews for each shortlisted candidates before selection."* – Engineering Lead, AXION Platform Project

In order to speed up this talent recruitment and management process, MTech setup a US subsidiary, MTech Imaging USA LLC, in 2009. Besides leveraging it as a talent recruitment center, the US subsidiary also provides direct access to a complement of experts that helped to develop the 'magical' advanced infrared imaging processing and application algorithms in the AXION platform.

*"Where I live in California, there are a number of infrared camera companies that make detectors, cameras and instruments … The University [in California] has a lot of programmes specific to thermal imaging. In that environment, we have quite a few people walking around that know a lot about infrared [technologies] with many years of experiences. Not just college's experiences, but they worked in two or three infrared companies with 10, 12 or 15 years of experiences. Once in a while, you would like to have these experienced guy to help you with your task and with your design [for the AXION platform]. It is more difficult to find that kind of people here [in Singapore]."* – Mr Bob Nishi, General Manager, MTech Imaging Pte Ltd

MTech's management resolved this talent challenge by implementing two strategies: (1) leveraging the Singapore brand name as a modern city nation and one of the safest and most stable financial hubs in the world to attract good applicants including those in China and India; and (2) creating a scheme that allowed thermal imaging experts especially those in USA to work as consultants.

After struggling for a few years with the new business model which included nine months of extensive search and recruitment activities, an international AXION team was finally formed. See Exhibit 8 for the AXION's research and development team structure (each functional area within a team is the responsibility of one person. Thus, the staff strength of the AXION team is 9).

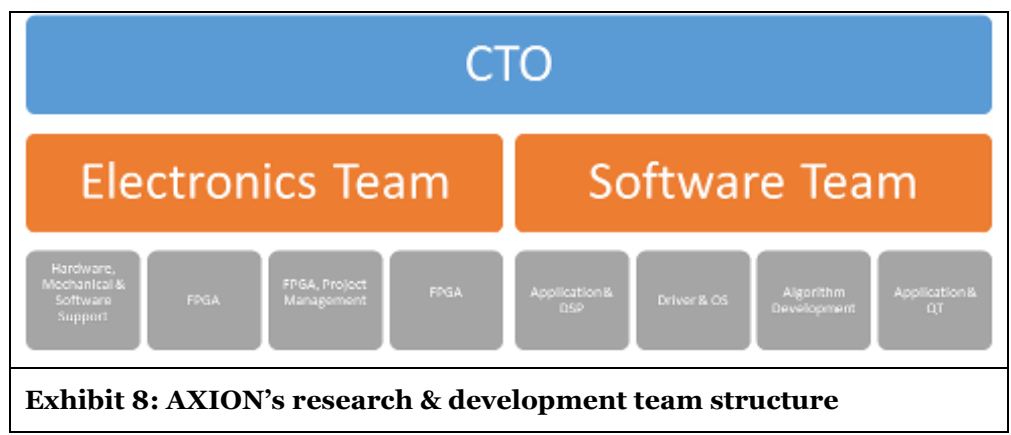

**Exhibit 8: AXION's research & development team structure**

### 5.2  Managing the AXION Team

To overcome challenges in managing a diverse team of different expertise and cultures, MTech's management undertook various important measures to ensure a cohesive and highly innovative culture.

First, time was dedicated for face-to-face working meetings in the initial months of the AXION project to develop an effective working culture and to inculcate strong camaraderie among team members.

Second, MTech's management empowered the CTO to make all critical decisions to drive the direction of the AXION project.

*"It was a CTO-run team. Everyone is either hardcore scientists or engineers … we only provide some support in [business] strategies during the project. We gave the CTO total leeway on how he wanted to run the project"* – Mr Baljit Singh, Executive Director, MTech Imaging Pte Ltd

Third, the project progress was closely monitored with at least one team meeting and a few of group discussions every week during the development phase. Each member was mandated to report his/her weekly activities and plan for the coming week to the project manager. This was then compiled and reported to MTech's management.

The meetings were also designed to create opportunities for members to exchange ideas and learn different areas of expertise from each other. Such cross-learning benefitted everyone as the members begun to develop in-depth understanding of how they can collaborate and synergize their efforts.

All progress and plans were systematically documented online via a document management system called Redmine which facilitated the dissemination of important information and allowed MTech's management to manage the team's development more effectively. The same system was also used by the team to share ideas, knowledge and information.

*"The [Redmine] platform was central to our success in managing knowledge and development experience of the team. The team members were very diligent in contributing to the retention, update and maintenance of information for any issue pertaining to the product development. The information contributed included data from external sources. Any team member could have access to the platform and contribute."* – Engineering Lead, AXION Project

## 5.3  Nurturing MTech's Core Competencies

To manage the inherent uncertainty in the AXION project, the team invested a significant amount of time and efforts on preparing, performing, and reviewing the identified critical tasks of the project. These reviews aided the team to establish more realistic project schedule with tighter estimates of the time and effort needed to complete variety of uncertain tasks.

All identified complex tasks were systematically broken down into multiple smaller tasks and assigned to different team members. To ensure alignment, members were required to periodically check and validate each other's progress until the complex task was completed.

To balance between time-to-market and quality, MTech's management adopted a rapid prototyping development approach. In each of the identified AXION product life-cycle plan, a basic version would be introduced to solicit valuable feedback from customers for features to be included in subsequent releases. The baseline version will be progressively updated with features and software libraries that better meet customers' expectations (see Exhibit 9 for details of AXION Development Roadmap).

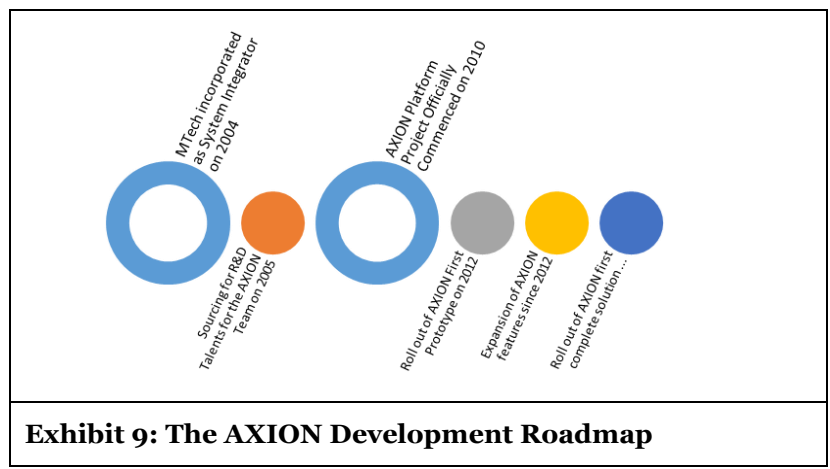

**Exhibit 9: The AXION Development Roadmap**

*"We continue to add more new capabilities to AXION and some of these makes the AXION even more capable than originally envisaged. We believe it is positioned as a market leader in terms of some of the latest features that we have added.  We recently added GPS, Compass, Gyro, Accelerometer, and WiFi streaming to the AXION [platform].  We have plans in the near future to add AXION compatibility to other thermal cameras and this will further position AXION as the smart engine for two of the world's largest thermal sensor manufacturers."* – Engineering Lead, AXION Project

When the AXION product was first rolled out to customers, MTech's management was surprised and disappointed by the initial response.

*"As good and sound a business plan that we put up, the market actually made us changed our business plans quite quickly when we went out to test it … we tested the market trying to sell just the AXION platform [without the camera] … but the market's request was they wanted a smart camera platform [which include the camera and the AXION platform] … so we had to make a massive shift*

*[in our business strategies] ... Interestingly, this makes us more confident with our products"* Mr Baljit Singh, Executive Director, MTech Imaging Pte Ltd

The team quickly adopted a camera developed by the market incumbent (FLIR) into the platform and package it as a complete solution to customers with greater reception. The success of the AXION platform development was not just solely the efforts of MTech's management and its AXION team. It also required MTech to work closely with reliable PCB manufacturers that built the hardware. The compactness and versatility of the AXION platform design made the assembly process extremely challenging. Furthermore, most PCB contract manufacturers in Singapore catered only to large volume production rather than smaller quantities needed by MTech for its prototyping approach.

## 6    The Management Challenge

Richard reflected upon the challenges and proposals that were discussed in MTech's meeting room on how to drive the market adoption of its AXION platform. First, there were regulatory restrictions on the export of US infrared technologies to countries like China and India especially in the defence industry. This was seemingly trivial to address given the programmability and flexible feature of the AXION platform.

*"Using India as an example. The country is very sceptical about the use of US full-frame camera [to detect infrared], so when you tell them that you can just remove the camera and replace it with a France-made camera, they are happy"* – Mr Baljit Singh, Executive Director, MTech Imaging Pte Ltd

Second, adoption of AXION platform in a specific industry will require extensive subject matter expertise. For instance, integrating the AXION platform into a medical device will need consultation with doctors to understand how infrared technology is able to solve medical challenges. It is financially infeasible for MTech to acquire such expertise. Hence, the idea of crowdsourcing was suggested. Akin to a users' toolkit as described by Hippel and Katz (2002), MTech is considering releasing its AXION platform as a user toolkit for software companies or experts in the thermal imaging field to develop unique solutions for a variety of problems in its respective domain. Once developed, these solutions will immediately be available to all customers which Richard believed will help to accelerate AXION's adoption. But how to do this in a scalable way without alerting its competitors prematurely? Does MTech have the capabilities to create a successful users' toolkits? What elements are needed to achieve success?

*"Moving forward, the company would engage software consultants to increase the number of software capabilities into AXION. This strategy to outsource future enhances our software capabilities and this helps to reduce the development cost."* – Engineering Lead, AXION Project

Third, solutions developed on the AXION platform could possibly service large number of markets. These may include security, law enforcement, firefighting, medical, thermography, automotive, maritime, unmanned aviation vehicles, and handheld infrared devices. Should MTech focus on all these markets at once or a select few? If the latter, which markets would be most cost-effective for them?

Fourth, MTech can consider partnership with existing market leaders such as FLIR to tap on their capabilities and distribution network to rapidly grow AXION platform adoption. If this is viable, how should they approach FLIR? What are the advantages and disadvantages of this approach?

Fifth, MTech can remain independent by leveraging its existing distribution network. While this may result in a slower pace of AXION platform adoption, it accords greater control. But how should they do it?

*"How can we bring our technology [AXION] to our customers in the fastest and most cost-effective way? I think the solution may be in distribution. Should we go direct selling? Or should we train value-added retailers or resellers who already in these market spaces? If yes, how should we up sell or down sell to them? This could be the challenge we face and we haven't figure it out ourselves yet."* – Mr Richard Yong, Chief Operating Officer, Miltrade Holdings Pte Ltd

Notwithstanding all these options have already been explored and discussed, Richard wondered if there are other better solutions to this management challenge.

For now, the search for the 'eureka' moment continues ...

# 7 Supplementary References & Readings for Students

## Copyright

# Visualizing the invisible – the relentless pursuit of MTech Imaging

Teaching Case Notes

## Case Synopsis

This teaching case describes a management challenge faced by MTech Imaging Private Limited (MTech), a Singapore-incorporated small & medium enterprise (SME) that specializes in thermal imaging since 2004. In more recent years, the company has relentlessly pushed itself to become a technology and innovative solutions provider in the thermal imaging industry. This relentless push by the company has paid off and has led to the development of a revolutionary innovation, the AXION platform, which the company believes that it will disrupt the industry. This case provides a rich contextual description of the journey undertaken by MTech to develop the AXION platform and provide a robust and somewhat conflicting discussions among senior managers of MTech on what would be their next strategic move.

To facilitate the initial discussion of the teaching case among lecturers and students, a summary of the history of MTech and its parent company, Miltrade Technologies Private Limited was presented. The case also showcases one of the key highlights of MTech's history, i.e. its contributions in helping Singapore's Government combated SARS in 2003. A brief description of the thermal imaging industry is also provided within the case to help students appreciate the environment that MTech competes in.

The senior management of the company believes that the AXION platform will not only potentially disrupt the entire thermal imaging industry, but will allow the company to emerge as a new industry's leader. Unfortunately, given the limited resources available as a SME, MTech has to determine strategically what would be the most effective move to maximize the benefits that can be derived from this disruptive innovation. Using the development journey of AXION as a leading story, the case provides a robust discussion on the current organizational capabilities of MTech especially in MTech's abilities to attract, manage and retain talents in the field of thermal imaging and directing these talents towards innovation development.

The case presents Mr Richard Yong, Miltrade's Chief Operating Officer as the protagonist and attempts to create an immersive storyline to make students feel the frustration and dilemma that Richard faces in strategizing what would be the next step for the company.

Students are expected to participate in two forms of role playing in class. The first is a team of disruptive technology's consultants who is required to submit to Richard a viable business proposal on how to drive the AXION platform (a disruptive technology). The second is a team of technical sales employed by MTech Imaging to sell the AXION platform to potential customers. The teaching case is designed to help students acquire the knowledge and skills needed to become a strong technology innovator.

## Target Audience

This teaching case is specially designed to provide a rich context that illustrates the challenges faced by SME in deriving disruptive digital innovations in the thermo-imaging industry.

It is specifically written for senior IS students who should be in its 3rd and 4th year of study. We highly recommend that the teaching case be used in one or more of the modules (or similar modules in nature) that are proposed in the recent paper published in MIS Quarterly by Fichman et al. (2014) on the "Information Technology in Business: A Digital Innovation Perspective" Programme. We share similar passion as Fichman et al. (2014) in pushing for a digital innovation curriculum in our School. The five modules as proposed by Fichman et al.

(2014) are listed below for your easy reference. Lecturer is strong encouraged to read their paper to have a more in-depth understanding of their proposal.

**(1) Module 1: Fundamentals of Digital Innovation**
*Topics*: Introduction to digital innovation, distinctive IT characteristics, process/product/business model innovation and cycles of digital innovation

**(2) Module 2: Digital Innovation at the Discovery Stage**
*Topics*: Identification of innovation opportunities, invention and selection.

**(3) Module 3: Digital Innovation at the Development Stage**
*Topics*: Developing ideas into usable innovations, packaging versus configuration

**(4) Module 4: Digital Innovation at the Diffusion Stage**
*Topics*: Deployment and assimilation of innovation

**(5) Module 5: Digital Innovation at the Impact Stage**
*Topics*: Value appropriation and transformation

We believe our teaching case is highly suitable to be used in Module 4 and/or 5 (or any modules of similar nature in your university). This is because we believe the case is open ended and contain good amount of information to provide a great opportunity for senior IS students to hone their knowledge application and presentation skills. The teaching case will demand students to articulate an effective strategy to help the protagonist of this case to deploy the digital innovation into the market so as to appropriate its values and to disrupt the industry.

## Learning Outcomes

The learning outcomes for this teaching case can be divided into two key components namely, theory and practical. They are illustrated as below:

Knowledge & Understanding (Theory Component)

1. Analyze the internal and external factors that are affecting the profitability and survivability of a SME using various strategy's analysis tools. This may include tools like the IS strategy triangle, the Porter's five forces, resource-based view, the PESTL analysis and disruptive innovation.
2. Formulate a suitable IS and business strategy to address the factors identified in (1) in order to achieve and sustain competitive advantage in the market.

Key Skills (Practical Component)

1. Discuss and defend the best IS strategy's recommendations using case information, facts and informed judgment.
2. Formulate convincing viewpoints during discussions that demonstrate strong critical thinking skill.
3. Demonstrate abilities to source insights from multiple information sources and combine them in a logical way to support their recommendations or sale pitch.

The teaching case emphasizes on the student's demonstration of skills in applying their knowledge in strategy formulation and in arguing and justifying their recommendations and sale pitch.

There are no right or wrong answer in each of the role playing assignment, but the most valuable learning experiences lie in their analysis of the situations and how they communicate and convince others about their recommendations or sale pitch.

## The Student's Assignment

The descriptions listed in this section should be disseminated to students in the early part of the semester so that they will have a good idea the assignment that they are supposed to work on using this teaching case

## Role-play Assignment 1 – Disruptive Digital Innovation's Consultant Team

Richard has decided to consult your company to develop the next course of action for MTech. You and your team are experts in driving disruptive digital innovations and have been assigned by your company to this project. Given the urgency to determine the strategy to drive its AXION platform, Richard wants a comprehensive business proposal within one week. He mandated that your proposal must consist of the following content (but not limited to):

1. Comprehensive market research and analysis
   a. Sufficient coverage of environment factors and/or enabling & competitive factors;
   b. Correlation of identified factors to provide useful insights
   c. Identification of potential gaps/opportunities how information resources can be used effectively to modify these environment factors to the favor of MTech.
2. Coherent strategies and programmes for disrupting the market using AXION platform
   a. Compelling go-to-market ideas supported with sound market analysis
   b. Creative use of online medias & channels to market AXION platform with the intent to disrupt the market eventually
   c. Differentiated and innovative propositions for its potential customers
   d. The way in which the platform will be governed by MTech Imaging
3. Feasibility of implementation plan
   a. Practical operational & roll-out plans
   b. The processes that need to be put in place to maintain the good relationships among the external developers/contributors of application for the AXION platform, MTech Imaging and its customers
   c. The intellectual property rights for external developers/contributors of applications for the AXION platform and the way it can be protected
   d. The way in which incentives can be incorporated into the AXION platform to allow maximum commitment from the external developers/contributors of application to the AXION platform and for MTech Imaging to monetize this platform successfully
   e. Adequate risk assessment and mitigation plan with contingencies

## Role-pay Assignment 2 – MTech Imaging's Technical Sales & Marketing Team

Shortly after the rolled out of the AXION platform, Richard setup a technical sales team to market to potential customers. You are one of the team members hired. Richard has the following potential customers that he would like you to pitch the AXION platform to.

1. **JZ Agri Products** is a major crop grower in the farming and agricultural industry. It is currently hiring helicopter services to perform monthly aerial surveillance on crop conditions. This has a very high operational cost and dependence on availability of aircraft and qualified pilots. Additionally, crop surveys are subjected to uncontrollable weather conditions and limited by guidelines on aircraft type, pilot flight duration and rest hours amongst other things. The images captured using current method is unable to measure Crop Water Stress Index (CSWI), the relative transpiration rate occurring in plants. When a plant is transpiring fully the leaf temperature is 1 to 4 degrees below air temperature and CWSI is low. As the crop undergoes water stress when the transpiration decreases, the leaf temperature rises and can reach 4 to 6 degrees above air temperature. When the plant is no longer transpiring the CWSI is high.

2. **HomeSmart** is an equipment manufacturer and solutions provider specializing in home automation systems. It is planning to develop and launch a new product line for remote safety monitoring of elderly and handicapped. This shall be targeted at healthcare industry especially hospitals and nursing homes where constant attention by care providers may not be available. Current CCTV technologies are very dependent on good lighting conditions. At low lighting conditions, existing systems often need to be coupled with motion detectors, which are not effective for detecting hazards or emergencies. Additionally, privacy concerns limit customer acceptance because a person's identity and detailed actions could be viewed or recorded. Motion detectors are often unable to provide accurate indication of actual situation. For instance it cannot differentiate whether the motion is caused by surrounding environment or human accident or other health situations. The rate of false alarm is high and therefore presents an opportunity for better solutions.

HomeSmart is exploring a thermal imaging solution and uncertain about costs, functions, quality and pricing model to be market viable. It considers purchase in excess of 500 units per customer as a reasonable volume to base their budget on.

3. **PetroViz** is an MNC that owns a few plants in Singapore, including petrochemical and power. Its current plant security and surveillance system is dependent on human to perform foot patrol around its vast operating area with CCTV monitoring its perimeters. To 'see' in the dark, all secured areas need to be brightly lit at night to be able to detect anything amiss. Additionally, the CCTV is unable to differentiate the source of movement or motion e.g. whether they are from animals or actual intrusion, resulting in high false alarm rate. Hence, PetroViz is exploring alternative technology that could provide wireless videos on surveillance and unmanned version could be an option. Although it's required quantity is not high but it is willing to pay premium for proven and reliable systems/technologies that will work for their plants and possibly those in other territories.

You will lead your team to make an important sales pitch to the senior management of one of the above customers. Richard will inform one week before the scheduled meeting which customer you will be presenting to. You are expected to make a convincing presentation and secure the sales of AXION platform.

## Teaching Plan

In order to achieve the above mentioned learning outcomes, we have designed a 'management game' based on this teaching case. We have designed the teaching case to be discussed in a team of four or five. The case is ideal for a class size of 25-30, which means that there should be around 6 teams. The management game can be broken down into four sequential activities that we suggested to be implemented at least a week apart from each other in a typical 12 weeks semester. In total, the students are expected to form a 4-5 persons team to perform the following tasks:

1. Develop and submit a business proposal (max of 20 pages including everything). The business proposal will be marked using the marking rubric indicated in Exhibit 1.

|  |  | Excellent | Good | Average | Fair | Poor |
|---|---|---|---|---|---|---|
| **Category** | **Weight** | **5** | **4** | **3** | **2** | **1** |
| Comprehensive Market Research & Analysis | 25% |  |  |  |  |  |
| Coherent Strategies & Programmes | 25% |  |  |  |  |  |
| Feasibility of Implementation Plan | 20% |  |  |  |  |  |
| Subtotal | 70% |  |  |  |  |  |
| **Exhibit 1: Marking Rubric for Business Proposal** | | | | | | |

2. Present their business proposal to the lecturer and their classmates (as senior management of MTech Imaging). The lecturer will score the business proposal using the marking rubric indicated in Exhibit 2.

|  |  | Excellent | Good | Average | Fair | Poor |
|---|---|---|---|---|---|---|
| **Category** | **Weight** | **5** | **4** | **3** | **2** | **1** |
| Logical flow & structure<br>- Well organized?<br>- Comprehensive? | 10% |  |  |  |  |  |
| Oral delivery<br>- Clear & concise<br>- Active engagement with audience<br>- Project & instil confidence | 5% |  |  |  |  |  |
| Visual presentation<br>- Good use of slides & props to | 5% |  |  |  |  |  |
| reinforce proposition | | | | | | |
| Q&A<br>- Demonstrate teamwork<br>- Good response to question | 10% | | | | | |
| Subtotal | 30% | | | | | |

**Exhibit 2: Marking Rubric for Business Proposal Presentation**

3. Develop and present their sale pitch to the lecturer and their classmates (as senior management of the customers) in class. The lecturer will score the sale pitch using the marking rubric indicated in Exhibit 3.

| | | Excellent | Good | Average | Fair | Poor |
|---|---|---|---|---|---|---|
| **Category** | **Weight** | **5** | **4** | **3** | **2** | **1** |
| Well-tailored presentation<br>- Effective use of time<br>- Concise & good communication skills exhibited | 30% | | | | | |
| Salesmanship<br>- Establish rapport & engage with customers actively | 20% | | | | | |
| Understanding of context<br>- Demonstrate good understanding of market, needs and situations of customers to provide suitable recommendations | 30% | | | | | |
| Q&A<br>- Attentive to prospect's priorities & response confidently<br>- Able to spot time to close sale | 20% | | | | | |
| Total | 100% | | | | | |

**Exhibit 3: Marking Rubric for Sale Pitch Role Play**

We recommend that the overall distribution of the weights of the scores for all three activities when computing the final score of each team.

- **Activity 1 & 2 (70%) + Activity 3 (30%) = 100%**

The detailed of the teaching and learning activities for this teaching case are listed below (note that the activities listed below are designed for a class size of 25-30 and should be adjusted accordingly if the class size differs from this).

1. **Briefing and Team Formulation (3 hours)** – This should be done in the first week of class for the course, where possible. In this briefing, the rules of the game, the expectation of students, and a brief introduction of the teaching case will be provided in this activity.

   After this is done, we recommend the lecturer spends some time to provide a quick summary of all the IS strategy related concepts and tools. This summary can include the following:
   a. Types of Information Resource (Piccoli and Ives 2003);
   b. Porter's five forces and how IS resources can be used to change them (Pearlson and Saunders 2013);
   c. Resource-based theory and what are the IS resources and how they can be used strategically to achieve and sustain competitive advantages (Wade and Hulland 2004);
   d. IS strategy triangle (Pearlson and Saunders 2013); and
   e. Disruptive innovation (Christensen and Overdorf 2000). We recommend lecturer considers running a video on Clayton's disruptive innovation as shown in Exhibit 4.

   This will help the students appreciate how disruptive innovation can disrupt an industry.

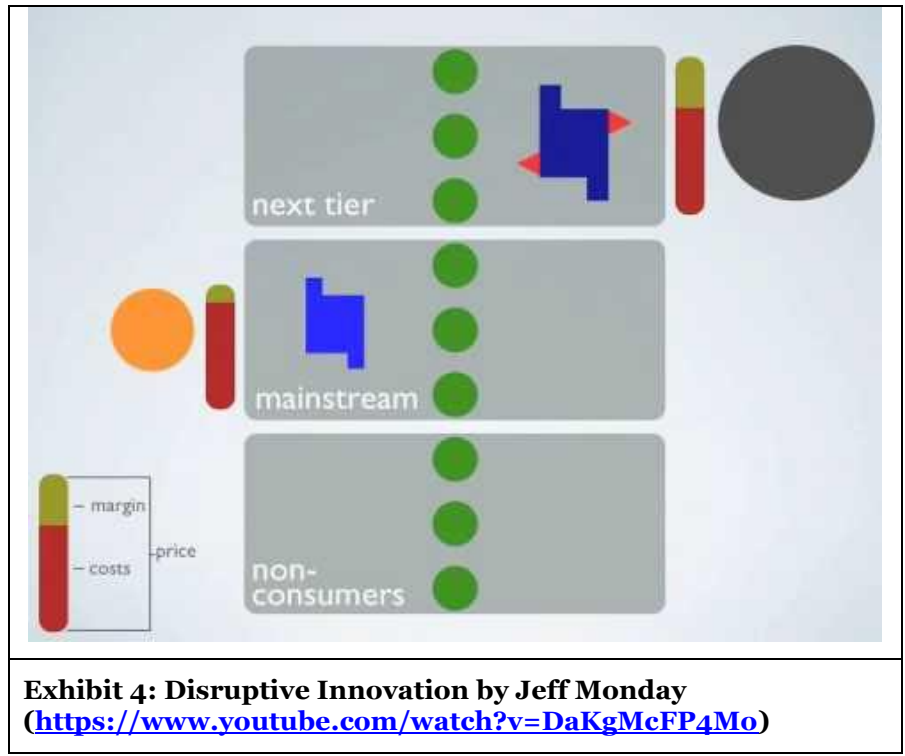

**Exhibit 4: Disruptive Innovation by Jeff Monday (https://www.youtube.com/watch?v=DaKgMcFP4Mo)**

2. **Role Play 1 Presentation (3 hours)** – This activity should be held at least one week <u>after the briefing activity</u>. Teams are expected to have submitted their business proposal and lecturer is expected to have graded their proposals before this presentation. Each team will be given 30 minutes (20 minutes presentation and 10 minutes Q&A) to present their proposal in class. They are to treat everyone in the class including the lecturer as part of the senior management of MTech Imaging. Lecturer will mark the presentation using the marking rubric as indicated in Exhibit 2.

   At the end of this activity, lecturer should provide a summary of the strengths and weaknesses of each presentation to reinforce students' learning.

3. **Role play 2 Presentation (3 hours)** – This activity is to be held at least one week <u>after the role play 1 presentation activity</u>. In this activity, teams will be asked to switch their roles from being consultants to being employees of MTech. *<u>A potential customer profile will be randomly assigned to the team one week before this activity by lecturer</u>*.

   The team is expected to make strong attempt to sell the AXION Smart Infrared product to the assigned company. Teams are asked to come up with a set of slides that will be used to pitch MTech's products to the customer. The slides are to be submitted to the lecturer two days before the scheduled class. The lecturer will play the role of the senior management of the potential customer and will grade the sale pitch presentation using the marking rubric as indicated in Exhibit 3.

   It is important during this stage that lecturer tries to play the role of 'devil advocate' to question the feasibility and benefits of AXION. Each team will be given 20 minutes (10 minutes to present and 10 minutes of Q&A). Lecturer should round up all the good and bad practices after this activity to reinforce the learning of students.

4. **Post Game Summary (3 hours)** – This activity is to be held at least one week <u>after all the above mentioned activities</u>. To reinforce student's learning, lecturer should highlight the strengths and weaknesses of: (1) each proposal made by the team; (2) each role play 1 presentation; (3) each role play 2 presentation. The lecturer should also take this opportunity to single out some outstanding proposals/presentations and provide feedback to students on how they can further improve their skills in these areas.

An illustration of the sequence of activities in the teaching plan is presented in Exhibit 5.

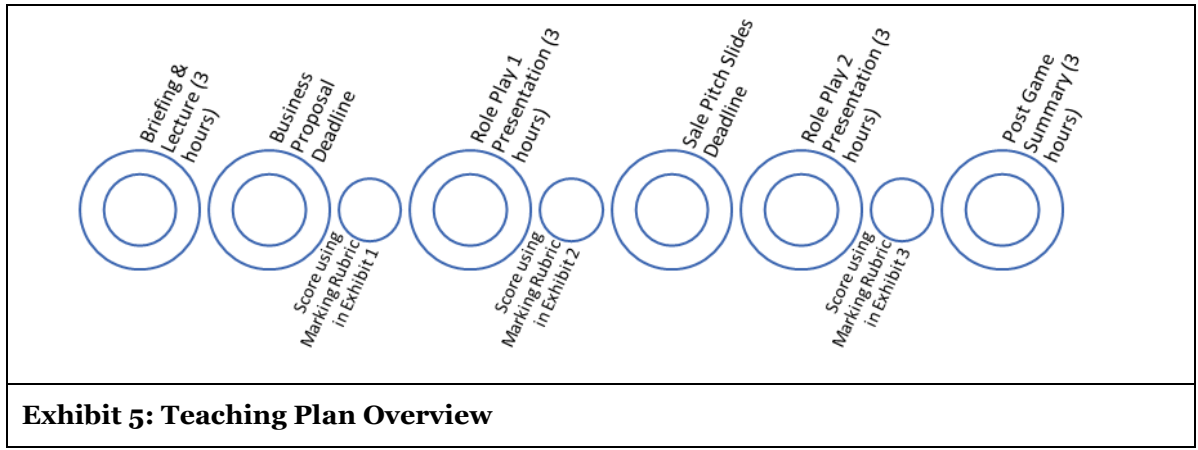

**Exhibit 5: Teaching Plan Overview**

## Supplementary Readings for Students

Christensen, C.M., and Overdorf, M. 2000. "Meeting the Challenge of Disruptive Change," *Harvard Business Review* (78:2), pp. 67-75.

Fichman, R.G., Dos Santos, B.L., and Zheng, Z.E. 2014. "Digital Innovation as a Fundamental and Powerful Concept in the Information Systems Curriculum," *MIS Quarterly* (38:2), pp. 329-353.

Gawer, A., and Cusumano, M. 2012. "How Companies Become Platform Leaders," *MIT Sloan Management Review* (49:2).

von Hippel, E., and Katz, R. 2002. "Shifting Innovation to Users Via Toolkits," Management Science (48:7), pp. 821-833.

Pearlson, K.E., and Saunders, C.S. 2013. *Strategic Management of Information Systems (5th Edition) - International Student Version*. John Wiley & Sons Singapore Pte. Ltd.

Piccoli, G., and Ives, B. 2003. "IT-Dependent Strategic Initiatives and Sustained Competitive Advantage: A Review and Synthesis of the Literature," *MIS Quarterly* (29:4), p. 755.

## Copyright

The following copyright paragraph must be appended to the paper. Author names should not be included until after reviewing. Please ensure the hyperlink remains for electronic harvesting of copyright restrictions.